\documentclass[twocolumn,amsmath,amssymb]{revtex4}

\usepackage{graphicx}

\begin{document}

\title{Closed-form solutions and scaling laws for Kerr frequency combs}

\author{William H. Renninger and Peter T. Rakich}
\affiliation{Department of Applied Physics, Yale University, New Haven, Connecticut 06520}

\begin{abstract}
A single closed-form analytical solution of the driven nonlinear Schr\"{o}dinger equation is developed, reproducing a large class of the behaviors in Kerr-comb systems, including bright-solitons, dark-solitons, and a large class of periodic wavetrains. From this analytical framework, a Kerr-comb area theorem and a pump-detuning relation are developed, providing new insights  into soliton- and wavetrain-based combs along with concrete design guidelines for both. This new area theorem reveals significant deviation from the conventional soliton area theorem, which is crucial to understanding cavity solitons in certain limits. Moreover, these closed-form solutions represent the first step towards an analytical framework for wavetrain formation, and reveal new parameter regimes for enhanced Kerr-comb performance.
\end{abstract}

\maketitle
Through Kerr-comb generation, octave-spanning optical frequency combs and high repetition rate (GHz-THz) pulse-trains can be generated by injecting continuous-wave laser-light into high quality factor microresonators \cite{Del'Haye2007,Savchenkov2008,Grudinin2009,Levy2009,Razzari2009,Liang2011,Kippenberg2011,DelHaye2011,Okawachi2011,Ferdous2011,Okawachi2014}.
Numerous studies have shown that cascaded parametric interactions within such cavity systems produce a tremendous variety of spectral features and intra-cavity waveforms \cite{Braje2009,Matsko2013,Herr2013,Leo2010,Godey2014}.  While wide-band spectral generation has been achieved in many systems \cite{DelHaye2011,Okawachi2011}, it is more challenging to identify the subset of conditions that permit stable and phase-coherent frequency comb formation. To elucidate the dynamics of comb formation, Kerr combs have been modeled with coupled-field \cite{Chembo2010,Chembo2010a,Hansson2013,Loh2014} and single-field \cite{Matsko2011,Coen2013a,Lamont2013,Coen2013,coillet2013,Parra-Rivas2014,Del'Haye2014,Godey2014} formulations of the damped-driven Nonlinear Schrodinger equation (NLSE), also termed the Lugiato-Lefever equation \cite{Lugiato1987,Coen2013a,Chembo2013}.  Numerical studies suggest that two archetypal patterns can be observed in different regimes of Kerr-comb operation: solitons and periodic wavetrain solutions (also termed Turing patterns and primary combs). Solitons have been observed through Kerr-comb experiments and described using perturbative analytical methods \cite{Leo2010,Herr2013}. In contrast, wavetrain-based frequency combs have only been observed numerically \cite{Godey2014}, and an analytical framework for wavetrain physics is lacking. To date, a universal and simple analytical framework that captures the diversity of nonlinear phenomena within Kerr-comb systems remains elusive.

Here we develop a single closed-form analytic solution of the driven NLSE that, remarkably, reproduces the behavior of both soliton and wavetrain based Kerr frequency combs.  This solution is validated with full microresonator numerical simulations, longitudinally invariant simulations, and established experimental results.  Applying this analytical framework, we derive a modified area theorem (or Kerr-comb area theorem) that relates the pulse duration to the peak powers within Kerr-comb systems for both soliton and wavetrain solutions, revealing significant deviation from the conventional soliton area theorem and tremendous opportunity for wideband comb formation using wavetrains.  A pump-detuning relation identifies conditions under which the corresponding waveforms are generated. Together, these relationships provide a design guide and predict new high performance regimes of broadband microcomb operation.  These results are generally relevant to microresonators, fiber Kerr-combs and Kerr-based systems pumped with a continuous-wave source.

As the basis for comparison with analytical solutions, full single-field numerical simulations are developed.  The parameters used are taken from widely studied silicon nitride microresonator experiments \cite{Levy2009,Lamont2013}.  Pulse propagation inside the microring is governed by the damped-driven NLSE of the form
\begin{equation}
\begin{aligned}
\frac{\partial E}{\partial z}=-\frac{\alpha}{2}E-i\delta E+i\frac{\beta_2}{2}\frac{\partial^2 E}{\partial t^2}+i\gamma|E|^2E.
\end{aligned}
\label{eqinring}
\end{equation}
Here, $E$ is the slowly-varying electric field envelope, $z$ is the propagation coordinate, $t$ is the local time or fast time, $\alpha$ is the linear loss, $\delta$ is the pump detuning, $\beta_2$ is the group-velocity dispersion (GVD), and $\gamma$ is the nonlinear Kerr coefficient.  In experiment, pump detuning ($\delta$) represents the frequency deviation of the pump light from a cavity resonance.  This model is strictly valid for bandwidths supporting pulses 50fs and longer and is easily expanded to higher order dispersion and nonlinearities \cite{Lamont2013}. Output coupling and continuous-wave pumping are modeled as lumped effects at one point in the cavity.  The temporal window is determined by the round trip cavity propagation time, which is given by the microring radius and the effective index.  The model is solved with a standard split-step fourier transform technique with a variety of hard excitations, and the solutions are considered stable when the change in energy per round trip converges to a numerically limited value ($\Delta E\lesssim10^{-14}$).

\begin{figure}[htb]
\centerline{
\includegraphics[width=8.0cm]{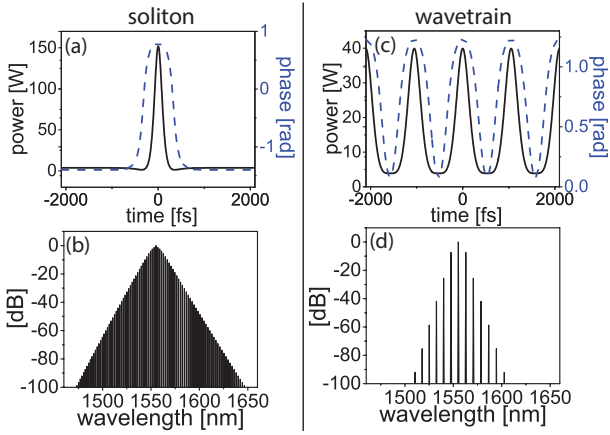}}
\caption{Numerical simulations of a 100-$\mu m$ radius SiN$_2$ microresonator with $\alpha=19.1 m^{-1}$, $\beta_2=2$ps$^2/$m, $\gamma=1/(Wm)$, and 0.25$\%$ output coupling. Soliton mode-locking temporal (a) intensity and phase profiles and (b) spectrum with $\delta=63.7m^{-1}$ and 5.4-mW input power (2.16-W pump).  Wavetrain mode-locking temporal (c) intensity and phase profiles and (d) spectrum with $\delta=0.16m^{-1}$ and 3.1-mW input power (1.24-W pump).}
\label{realsims}
\end{figure}

Of the many possible solutions, we focus on two solution types that are most pertinent to Kerr-comb formation: soliton solutions and wavetrain solutions \cite{Leo2010,Herr2013,Godey2014}.  The solitons (Fig. \ref{realsims}(a,b)) are characterized by a localized solitary wave atop a continuous-wave background.  A characteristic structure appears at the interface of the pulse with the background, which corresponds to a temporal phase shift.  Multiple solitons can also exist given a sufficient temporal window.  In the special case when a single soliton exists in the cavity, the free-spectral range (FSR) of the frequency comb is independent of the soliton and is given by the length of the microresonator.  Solitons exist with negative detuning ($\delta>0$).

Wavetrain solutions (Fig. \ref{realsims}(c,d)) are characterized by a periodic stable wave-form.  The waveform deviates from a sinusoid at the troughs of the intensity profile, which is also derived from a temporal phase change.  The FSR of these frequency combs is determined by the solution itself, although constrained to an integer multiple of the cavity round trip time. These solutions can exist when the cavity detuning is zero or positive as well as negative \cite{Godey2014}.  It should be noted that, like solitons, these wavetrain solutions are stable nonlinear attractors which possess an independent basin of attraction.  We will demonstrate in this work that all of these qualitative features are predicted by the analytical model presented here.

In order to study these classes of solutions, we look for their archetypal forms, as often exists in pattern forming systems of experimental relevance.  We begin with the master equation for the field,
\begin{equation}
\begin{aligned}
\frac{\partial E}{\partial z}=-\frac{\alpha}{2}E-i\delta E+&i\frac{\beta_2}{2}\frac{\partial^2 E}{\partial t^2}+i\gamma|E|^2E+ge^{i\psi},
\end{aligned}
\label{eqfull}
\end{equation}
which contains all of the relevant dynamics of the system. Note that this master equation is identical to Eq. \ref{eqinring}, but now $\alpha$ includes the output-coupling and the continuous-wave gain with arbitrary phase is distributed around the cavity.  The simplest system which exhibits the archetypal forms from above is the driven NLSE. This model has been shown to be accurate for systems with high quality factor and can often be extended to the case with larger loss with careful analysis \cite{Barashenkov1996}.  While relatively new in comparison to the undriven case \cite{whitham2011linear,novikov1984theory}, the driven NLSE has been shown to support a wealth of nontrivial phenomena \cite{Kaup1978,Nozaki1983,Nozaki1984,Nozaki1986,Friedland1998,Terrones1990,Haelterman1992}. Note that stable solutions to Eq. \ref{eqfull} exist without smallness requirements on the parameters \cite{Barashenkov1996}. An exact soliton solution is known for this driven NLSE \cite{Barashenkov1996}, and in 2005 \cite{Raju2005} a broader class of solutions was discovered; these solutions are of the form $E=\left(G+Kf(T)^2\right)/\left(1+Lf(T)^2\right)$, where $f$ are the Jacobi Elliptic functions.

We begin our analytical treatment of Kerr comb systems by expressing the driven NLSE, which governs Kerr-comb dynamics, in normalized form as
\begin{equation}
\begin{aligned}
\frac{\partial U}{\partial Z}=-is_{\delta}U+&s_{\beta}i\frac{\partial^2 U}{\partial T^2}+i2|U|^2U+he^{i\psi}.
\end{aligned}
\label{eqnorm}
\end{equation}
Here, $U=E\sqrt{\gamma/2|\delta|}$, $Z=z|\delta|$, $T=t\sqrt{2|\delta|/D}$, $h=\sqrt{g^2\gamma/2|\delta|^3}$, $s_\delta$ is $sign(\delta)$ and $s_\beta$ is $sign(\beta)$.  In these normalized variables, the pulse duration (defined as for the NLSE as the pulse duration parameter of a hyperbolic secant) and peak power can be written in real units as
\refstepcounter{equation}\label{peakanddur}
\[
P=\frac{2|\delta|}{\gamma}|U_0|^2,\quad\text{and}\quad\Delta\tau^2=\frac{|D|}{2|\delta|}\Delta T^2. \tag{\theequation\:a,b}
\]
Here, $|U_0|^2$ and $\Delta T^2$ are the unitless peak power and pulse duration and are determined by the solution to Eq. \ref{eqnorm}.  These equations can be written in a more instructive form as
\refstepcounter{equation}\label{eqAT}
\[
P\Delta\tau^2=\frac{|D|}{\gamma}F,\quad\text{and}\quad|\delta|=\frac{|D|\Delta T^2}{2\Delta \tau^2}=\frac{P\gamma}{2|U_0|^2}, \tag{\theequation\:a,b}
\]
where $F=|U_0|\Delta T^2$.  We term Eq. \ref{eqAT}a the Kerr-comb area theorem; it relates the pulse parameters to the system parameters in a general way.  Eq. \ref{eqAT}b is a complementary relation that determines the required detuning for a given comb duration or peak power.

For analysis of frequency combs, we further generalized the known class of solutions given in Ref. \cite{Raju2005} to include a phase factor, $\psi$, that accounts for the phase difference between the resonantly circulating cavity mode and the coherent pump (or drive). The relation of the phase on the pulse to that of the coherent pump is determined by the so-called ansatz technique. In addition, for convenient representation of both soliton and wavetrain forms we set $f=cn$.  Thus, the generalized closed-from analytic solution to Eq. \ref{eqnorm} is given by
\begin{equation}
\begin{aligned}
U=e^{i\psi+i\pi/2}\frac{G+Kcn(T/\tau,m)^2}{1+Lcn(T/\tau,m)^2},
\end{aligned}
\label{gensol}
\end{equation}
where $cn$ is the Jacobi Elliptic function and $G$, $K$, $L$, $\tau$ and $m$ are real pulse parameters which are determined by the system parameter, $h$. The pulse parameters are solved for by inserting the ansatz into Eq. \ref{eqnorm} with $\frac{\partial U}{\partial Z}=0$ and separately satisfying the real and imaginary parts. This process requires the solution of four nonlinear algebraic equations. Before examining the solution for arbitrary values of $m$, it is instructive to first examine the solution for the values where the $cn$ function reduces to the hyberbolic cosine ($m=1$) and the cosine ($m=0$) functions.

With the particular value of $m=1$, Eq. \ref{gensol} reduces the the soliton solution and can be written as
\begin{equation}
\begin{aligned}
U=e^{i\psi+i\pi/2}\left(CW+\frac{A}{cosh(T/\tau)+B}\right),
\end{aligned}
\label{coshsol}
\end{equation}
where $CW$, $A$, $B$, and $\tau$ are real parameters.    The four nonlinear algebraic equations, in this case, have two solutions. One solution is known to be unstable \cite{Barashenkov1996} and the other (stable) solution is given by
\begin{equation}
\begin{aligned}
&A=\frac{\sqrt{3}}{\tau  \sqrt{2 \tau ^2+1}},\quad h=\frac{\sqrt{\tau ^2-1} \left(2 \tau ^2+1\right)}{3 \sqrt{6} \tau ^3},\quad\\
&\text{CW}=-\frac{\sqrt{\tau ^2-1}}{\sqrt{6} \tau },\quad B=-\frac{\sqrt{2}
   \sqrt{\tau ^2-1}}{\sqrt{2 \tau ^2+1}}.
\end{aligned}
\label{coshrel}
\end{equation}
\begin{figure}[htb]
\centerline{
\includegraphics[width=8.0cm]{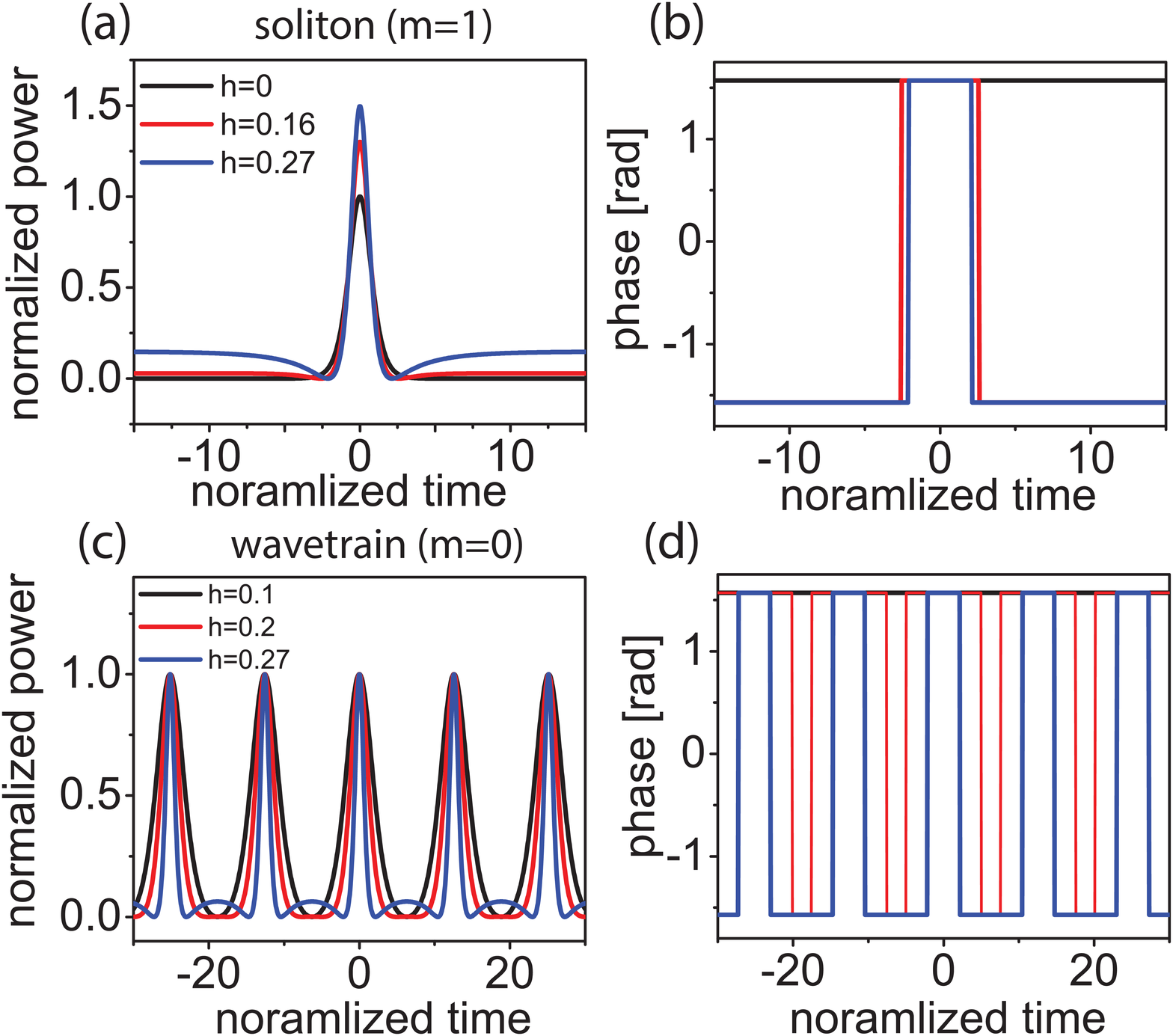}}
\caption{Representative temporal (left) intensity and (right) phase profiles for (a,b) soliton  (Eq. \ref{coshsol}) and (c,d) trigonometric wavetrain solutions (Eq. \ref{cossol}). The solitons are identified by $h$ including the $h=0$ solution and trigonometric solultions are also identified by $h$ but $h=0$ is not a solution. The cnoidal wavetrain solutions qualitatively the same as the trigonometric solutions.}
\label{ansatz}
\end{figure}
\noindent Here, the parameters are written in terms of $\tau$ instead of $h$ for compactness.  In addition, $s_{\delta}=1$. It is clear that when $h=0$ the equation reduces to the NLSE and the solution (Eq.  \ref{coshsol}) reduces to the well-known soliton solution of the form $Asech(AT)$. In this special case, $F =1$, and Eq. \ref{eqAT}a reduces to the familiar soliton area theorem. However, when $h>0$, the pulse develops a continuous-wave background with a characteristic dip in field amplitude at the wings of the pulse (Fig. \ref{ansatz}a).  This is the result of a phase shift when the electric field changes sign (Fig. \ref{ansatz}b).

\begin{figure}[htb]
\centerline{
\includegraphics[width=8.0cm]{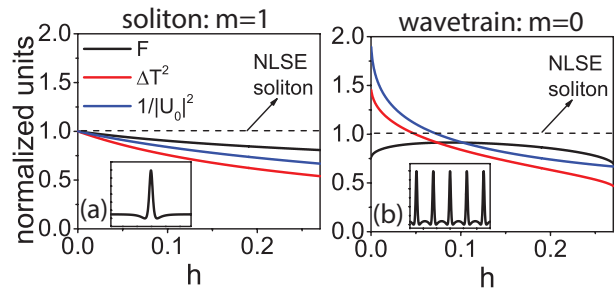}}
\caption{Kerr-comb area theorem prefactor ($F$), $1/|U_0|^2$ and $\Delta T^2$ for the (a) soliton and (b) trignometric wavetrain solution.  The dashed line represents the value for the NLSE soliton for comparison.  The pulse form is represented in the inset.}
\label{ATfig}
\end{figure}

$|U_0|^2$, $\Delta T^2$, $F$ are plotted for the soliton solution in Fig. \ref{ATfig}(a).  When $h=0$, $|U_0|^2$, $\Delta T^2$, $F$ are equal to 1 and Eq. \ref{eqAT} reduces to the well-known soliton area theorem and phase relation \cite{agrawal2001book}.  When $h$ takes on a maximum allowed value of $h=\sqrt{2/27}$, the prefactor of the Kerr comb area theorem becomes $F=0.8$, revealing significant deviation of the true cavity-soliton solution from the background-free NLSE soliton area theorem. Despite the clear differences in the solutions, the area theorem from the NLSE has theoretical justification for use to model experiments driven with a continuous-wave pump as was done in Ref. \cite{Coen2013}.  Through careful theoretical study, a subset of the solutions to Eq. \ref{eqnorm} were tied to the damped case (Eq. \ref{eqfull}) \cite{Barashenkov1996}, validating this solution as a general model for solitons in Kerr-comb systems.  It is also noteworthy that, in the normal dispersion regime, this same ansatz is a closed-form analytic solution for a dark soliton.

\begin{figure}[htb]
\centerline{
\includegraphics[width=8.0cm]{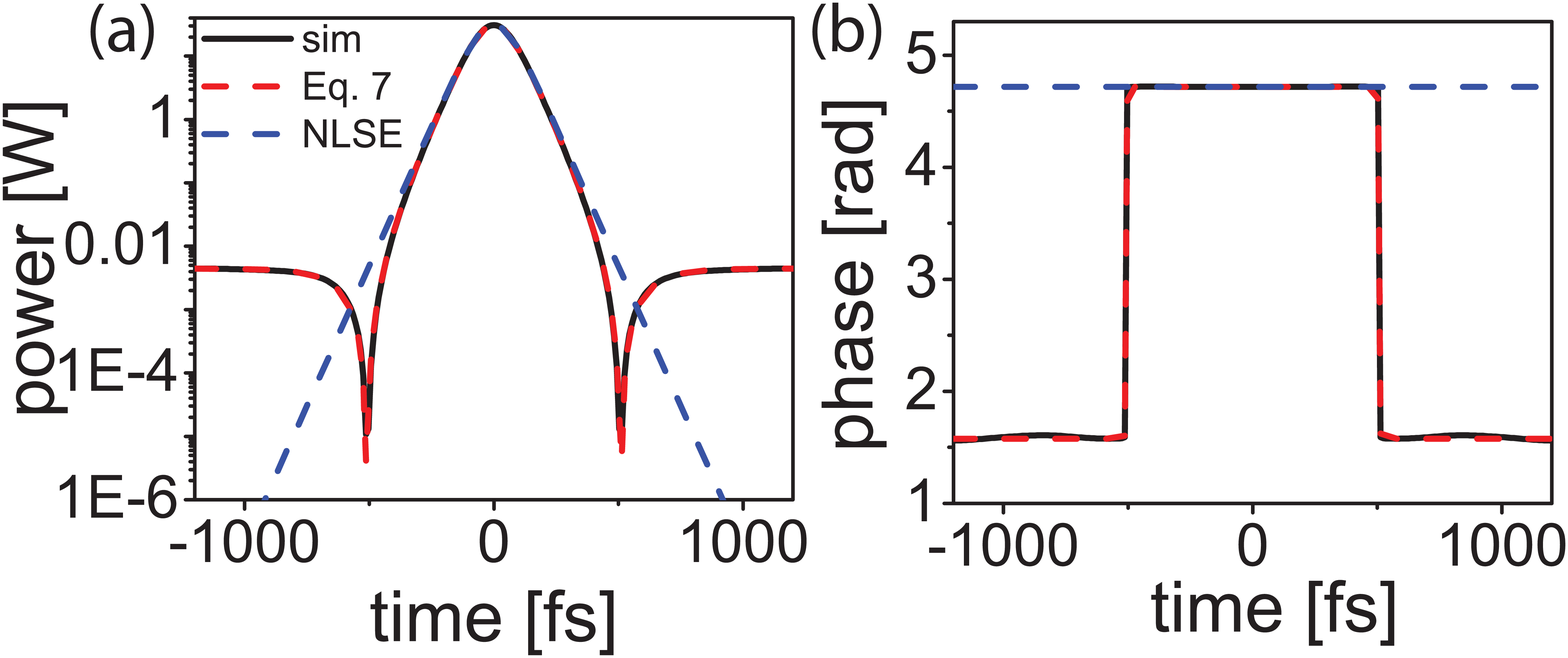}}
\caption{Numerical simualation of Eq. \ref{eqnorm}, with $\beta_2=0.3$ ps$^2/$m, $\gamma=1/(Wm)$, $\delta=15.0133/m$ and $g=0.35$ mW/m.  The temporal (a) intensity and (b) phase profiles are displayed (black solid) along with Eq. \ref{coshsol} (red) and the NLSE soliton solution with the same peak power (blue) for comparison.}
\label{compan}
\end{figure}

Numerical evaluation of Eq. \ref{eqnorm} produces convergence to a stable soliton solution, as seen in Fig. \ref{compan}. Despite the absence of a linear damping term in Eq. \ref{eqnorm}, energy decay occurs through spectral shedding; when convergence is achieved, the pump-wave no longer adds energy to the pulse due to wave-interference. For comparison, this numerical soliton solution of Eq. \ref{eqnorm} is plotted atop the analytic soliton solution (Eq. \ref{coshsol}),  revealing excellent agreement without any fitting parameters. The background-free NLSE soliton is also shown in Fig. \ref{compan}.  While the center of all three pulses show good agreement, divergence from the background-free NLSE solution is apparent when the phase changes and the continuous-wave background begins.

Another case of interest occurs when Eq. \ref{gensol} is evaluated for $m=1$. In this case, the solution can be written as
\begin{equation}
\begin{aligned}
U=e^{i\psi+i\pi/2}\left(CW+\frac{A}{cos(T/\tau)+B}\right).
\end{aligned}
\label{cossol}
\end{equation}
Hence, the solution exhibits a periodic time dependence.  In this case, the set of equations from which to the pulse parameters are found can be expressed as
\begin{equation}
\begin{aligned}
&A=\frac{\sqrt{3}}{\tau  \sqrt{2 \tau ^2-1}},\quad h=\frac{\sqrt{\tau ^2+1} \left(2 \tau ^2-1\right)}{3 \sqrt{6} \tau ^3},\quad\\
&\text{CW}=-\frac{\sqrt{\tau ^2+1}}{\sqrt{6} \tau },\quad B=\frac{\sqrt{2} \sqrt{\tau ^2+1}}{\sqrt{2
   \tau ^2-1}}.
\end{aligned}
\label{cosrel}
\end{equation}
This wavetrain solution is periodic with period $2\pi\tau$ (Fig. \ref{ansatz}(c)). The wavetrain deviates from a sinusoid at the troughs of the intensity profile as a result of a phase shift (Fig. \ref{ansatz}(d)).  The detuning for the trigonometric solution is also negative ($s_{\delta}=1$).  The Kerr-comb area theorem and phase relation can again be derived and written in the form of Eq. \ref{eqAT}. Figure \ref{ATfig}(b) reveals that the prefactor, $F$, is again close to 1.

\begin{figure}[htb]
\centerline{
\includegraphics[width=8.0cm]{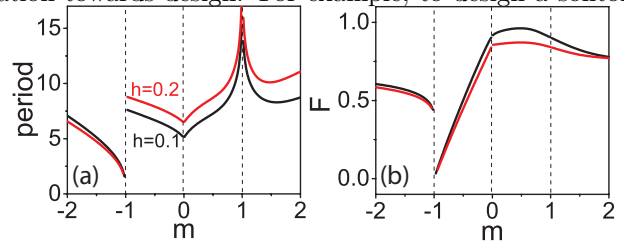}}
\caption{Trends of the cnoidal wavetrain solution.  (a) The wavetrain period vs. m and (b) Kerr-comb area theorem prefactor, F vs. m for two values of h.  The dashed lines represent discontinuous points in the roots to the nonlinear algebraic equations.  Note that the period approaches infinity at $m=1$, where the cnoidal solution reduces to the soliton solution. }
\label{cnsols}
\end{figure}

The broader class of solutions for arbitrary $m$ has pulse parameters which can be determined numerically from the solutions of the four nonlinear algebraic equations which relate them.  The period of the solution is given by $2\tau Re[K[m]]$, where $Re[K[m]]$ is the real part of the complete elliptic integral of the first kind (plotted vs. $m$ in Fig. \ref{cnsols}(a)).  As $m$ approaches 1 the period approaches $\infty$ where the localized soliton forms.  That is, when $m$ is very near one, Eq. \ref{gensol} represents an infinite train of soliton pulses of the form from Eq. \ref{coshsol}.  The broad class of wavetrain solutions are qualitatively similar to Eq. \ref{cossol}.  A Kerr-comb area theorem again takes the form of  Eq. \ref{eqAT}.  The prefactor, $F$, can be determined, as for the other cases, by solving four characteristic nonlinear algebraic equations.  The result is plotted as a function of $m$ for constant $h$ in Fig. \ref{cnsols}(b).
Interestingly there exists a regime of parameters where $F$ becomes much smaller than unity.  In this regime, for a transform-limited pulse, larger bandwidths can be achieved with a given peak power; this is a desirable trait for  wideband frequency comb applications.  Note that, in contrast to the bright soliton solution, which only exists with negative detuning ($\delta>0$) and anomalous dispersion ($\beta_2>0$), these periodic wavetrain solutions can exist at both signs of dispersion and detuning. This feature of the wavetrain solution is consistent with previous simulations and experiments \cite{Herr2013,Haelterman1992,Lamont2013}.  For the purposes of stable frequency comb generation these periodic solutions are intriguing; they permit tunability of the comb FSR, and a wider range of wave-forms over a parameter space that is much larger than soliton solutions.

The ability to relate the pulse parameters to the nonlinear system parameters (Eqs. \ref{eqAT}) enables direct application towards design.  For example, to design a soliton frequency comb with a given bandwidth ($\varpropto1/\Delta\tau$), Eq. \ref{eqAT}a can be used to determine the peak power, $P$, and Eq. \ref{eqAT}b can be used to determine the detuning, $\delta$.  Given the detuning, peak power and bandwidth we can calculate the required pump power for a given free-spectral range frequency comb. Hence, the parameters of the system are fully determined using this closed-form treatment.

While we have outlined the conditions for the existence of the cavity soliton and wavetrain solutions in Kerr-comb systems using this analytical framework, the stability of the solutions is paramount for their experimental observation. Wavetrain solutions have been studied extensively through  numerical simulations. While the existence of wavetrain solutions has been demonstrated in full simulations here (Fig. \ref{realsims}(c,d)) and in Refs. \cite{Haelterman1992,Lamont2013,Herr2013}, further study is required to rigorously connect the large class of wavetrain solutions produced by Eq. \ref{gensol} to the damped driven NLSE of Eq. \ref{eqfull}.  Hence, further numerical studies of the damped driven NLSE, with a focus on the wavetrain solutions, would be valuable.

By contrast, the stability of cavity solitons has been established through experiments \cite{Herr2013,Leo2010} as well as in full numerical simulations (Fig. \ref{realsims}(a,b)).  In addition, the analytical cavity soliton solutions presented here were shown to be stable with numerics (see Fig. \ref{compan}); following Ref. \cite{Barashenkov1996}, this class of soliton solutions is known to be stable with damping. The physical relevance of this analytical cavity soliton solution is corroborated by numerical models, analytical treatments, and recent experiments.

In conclusion, we have demonstrated a closed-form solution that represents a large range of Kerr frequency comb forms.  The cavity soliton is shown to deviate from the soliton solutions of the undriven NLSE, the wavetrains represent a stable nonlinear attracting solution with implications for high performance frequency combs, and a simple Kerr-comb area theorem and detuning relation were derived and used to generate experimental design guidelines and to predict new regions of performance.  These solutions provide an ideal ansatz for studies based on the variational approach for identifying new regimes of stable Kerr-comb generation.


\section*{Acknowledgements}
\noindent The authors acknowledge support from Yale University.

\end{document}